\documentclass[fleqn,usenatbib]{mnras}

\usepackage{newtxtext,newtxmath}
\usepackage[T1]{fontenc}

\DeclareRobustCommand{\VAN}[3]{#2}
\let\VANthebibliography\thebibliography
\def\thebibliography{\DeclareRobustCommand{\VAN}[3]{##3}\VANthebibliography}

\usepackage{graphicx}	% Including figure files
\usepackage{amsmath}	% Advanced maths commands
\usepackage{subcaption}

\defcitealias{2025MNRAS.539.3473Q}{QIAO25}

\title[Bolometric correction factor in TDEs]{Bolometric correction factor and radiative efficiency for the super-Eddington accretion flow in tidal disruption events}

\author[Yongxin Wu et al.]{
Yongxin Wu,$^{1,2}$
Erlin Qiao,$^{1,2}$\thanks{E-mail: qiaoel@nao.cas.cn}
Xuan Fang,$^{1,2}$
Yiyang Lin,$^{1,2}$
Jifeng Liu$^{1,2}$\thanks{E-mail: jfliu@nao.cas.cn}
and Meng Guo$^{3,4}$
\\
$^{1}$National Astronomical Observatories, Chinese Academy of
Sciences, Beijing 100101, China \\
$^{2}$School of Astronomy and
Space Sciences, University of Chinese Academy of Sciences, 19A Yuquan Road, Beijing 100049, China\\
$^{3}$Key Laboratory of Computing Power Network and Information Security, Ministry of Education, Shandong Computer 
Science Center (National \\
Supercomputing Center in Jinan), Qilu University of Technology (Shandong Academy of Sciences), Jinan, Shandong 250013, China\\
$^{4}$Jinan Institute of Supercomputing Technology, Jinan, Shandong 250103, China \\
}

\date{Accepted XXX. Received YYY; in original form ZZZ}

\pubyear{\the\year{}}

\begin{document}
\label{firstpage}
\pagerange{\pageref{firstpage}--\pageref{lastpage}}
\maketitle

% Abstract of the paper
\begin{abstract}
The estimate of the bolometric luminosity and the radiative efficiency are two key
aspects for understanding the properties of the accretion flow around a supermassive black hole (BH).
In this paper, we focus on the estimate of the bolometric luminosity and the radiative efficiency
of the early super-Eddington accretion flow in tidal disruption events (TDEs).
Specifically, we first perform radiation hydrodynamic simulations of super-Eddington accretion flow in TDE environment, and then calculate the corresponding emergent spectra with the method of post processing for the simulation data. Based on the emergent spectra, we calculate the isotropic-equivalent X-ray bolometric correction factor $k_\mathrm{bol}$ and the radiative efficiency $\eta$ of the super-Eddington accretion flow. We find that both $k_\mathrm{bol}$ and $\eta$ are BH mass and viewing-angle dependent. $k_\mathrm{bol}$
is in the range of about a few tens to a few thousands, and $\eta$ is in the 
range of $\sim 10^{-3}-10^{-1}$ for BH mass in the range of $10^{6-7}M_\odot$ and
the viewing angle in the range of $0^{\rm o}-90^{\rm o}$. Finally, we apply the derived 
$k_\mathrm{bol}$ and $\eta$ to some specific TDEs to estimate the accreted mass during an event, which can significantly alleviate the so-called missing energy problem in TDEs.

\end{abstract}


%%%%%%%%%%%%%%%%%%%%%%%%%%%%%%%%%%%%%%%%%%%%%%%%%%

%%%%%%%%%%%%%%%%% BODY OF PAPER %%%%%%%%%%%%%%%%%%

\section{Introduction}

When a star approaches a black hole (BH) within the tidal radius, the tidal forces from the BH exceed the star's self-gravity, leading to the disruption of the star, which is known as the tidal disruption events (TDEs) \citep{1975Natur.254..295H,1988Natur.333..523R}. 
Approximately half of the stellar debris on bound orbits will fall back and form an accretion disk through circularization by the self-intersecting collisions. 
The initial fallback rate is super-Eddington for BHs with masses between $10^6$ and $10^7M_\odot$ and decays with the form of $t^{-5/3}$ \citep{1989ApJ...346L..13E,1989IAUS..136..543P,2013ApJ...767...25G}, which leads to luminous radiation \citep{1990ApJ...351...38C,2021ARA&A..59...21G}.

X-ray TDEs were first detected in the 1990s by the ROSAT all-sky survey \citep{1996A&A...309L..35B,1999A&A...349L..45K,1999A&A...343..775K,1999A&A...350..805G}, which are characterized by radiation temperatures in the range of $\sim10^5$-$10^6$ K \citep{2021SSRv..217...18S}. It is widely accepted that X‑ray emission originates from accretion disks around supermassive BHs in super-Eddington states \citep{1999ApJ...514..180U,2026MNRAS.547ag496C}. TDEs have also been observed in the ultraviolet (UV) and optical band \citep{2006ApJ...653L..25G,2008ApJ...676..944G,2008ApJ...678L..13K,2009ApJ...698.1367G,2011ApJ...741...73V,2012Natur.485..217G,2014ApJ...793...38A,2014MNRAS.445.3263H}, which is thought to originate either from shocks produced by self‑intersection of the debris stream \citep{2015ApJ...806..164P,2016ApJ...830..125J,2020ApJ...904...73R,2024Natur.625..463S,2024ApJ...974..165H}, or from the reprocessing of accretion generated X‑ray emission into the optical band by optically thick outflows \citep{1997ApJ...489..573L,2009MNRAS.400.2070S,2016MNRAS.461..948M,2016ApJ...827....3R,2018ApJ...855...54R,2018ApJ...859L..20D,2022ApJ...937L..28T,2022MNRAS.516.2833B,2025MNRAS.540.3069P,2025MNRAS.539.3473Q,2026ApJ...998..193G}.

For X-ray selected TDEs, one can derive the total released X-ray energy by integrating the X-ray luminosity over time. An X-ray bolometric correction factor $k_\mathrm{bol}$ (defined as the ratio of bolometric luminosity to X-ray luminosity) is often adopted to correct X-ray to bolometric luminosity. The accreted mass during the outburst can be calculated as follows \citep{2021SSRv..217...18S}, 
\begin{equation}\label{eq1}
    \Delta M = \frac{k_\mathrm{bol}}{\eta c^2}\int_{t}^{\infty}L_\mathrm{X}(t)dt, 
\end{equation}
where $\eta$ is the radiative efficiency, $L_\mathrm{X}(t)$ is the X-ray luminosity at some time $t$ and $c$ is the light speed.

In \citet{2002ApJ...576..753L}, the authors integrated the soft X-ray light curve of NGC 5905 from 1990.54 to 1996.89 yr, yielding the total released X-ray energy, $\Delta E_\mathrm{X}=\int_{t}^{\infty}L_\mathrm{X}(t)dt\approx 4.5\times10^{49}\ \mathrm{erg}$. They performed blackbody fitting to the X-ray spectrum, using a constant X-ray bolometric correction factor, i.e., $k_\mathrm{bol}=1.1$ to calculate the total released energy $\Delta E$. They calculated the total accreted mass $\Delta M = \Delta E/\eta c^2 \approx \Delta E_\mathrm{X}/ \eta c^2\approx 2.5\times10^{-4}M_\odot (\eta/0.1)^{-1}$. If one takes a typical value of $\eta=0.1$, it gives $\Delta M=2.5\times10^{-4}M_\odot$. This value is significantly lower than the expected returning mass ($\sim0.5M_\odot$) from the disruption of a solar-type main sequence star. They pointed out that the derived lower value of $\Delta M$ is probably related to the underestimate to $k_\mathrm{bol}$.

The same method has been widely adopted to estimate $\Delta M$ for X-ray TDEs by assuming $k_\mathrm{bol}=1$ and $\eta=0.1$. For two X-ray TDE candidates NGC 3599 and SDSS J132341.97+482701.3, the total released X-ray energy $\Delta E_\mathrm{X}$ during the outburst is about $7.1\times10^{48}\ \mathrm{erg}$ and $7.6\times10^{50}\ \mathrm{erg}$ respectively, with which the total accreted mass is calculated to be $\Delta M \sim 4.0\times 10^{-5}M_\odot$ and $\Delta M \sim 4.2\times 10^{-3}M_\odot$ respectively \citep{2008A&A...489..543E}. For a large sample of X-ray TDEs, the radiated energy $\Delta E_\mathrm{X}$ typically ranges from $10^{49}$ to $10^{51}\ \mathrm{erg}$, corresponding to the accreted mass $\Delta M$ of approximately $10^{-4}$ to $10^{-2}M_\odot$ \citep{2009A&A...495L...9C,2010ApJ...722.1035M,2017A&A...598A..29S}. In the case of ASASSN-14li, comparable levels of X-ray and optical radiation were observed. The total radiative energy $\Delta E$ was estimated to be about twice of $\Delta E_\mathrm{X}$ , with a value of approximately $7\times10^{50}\ \mathrm{erg}$, which corresponded to $\Delta M\approx4\times10^{-3}M_\odot$ by assuming $\eta=0.1$ \citep{2016MNRAS.455.2918H,2017MNRAS.466.4904B}. The substantial discrepancy between  {the calculated} $\Delta M$ from observations and the theoretically expected value of disruption a solar-type main sequence star leads to the so-called `missing energy' problem.

Several solutions have been proposed to answer the ‘missing energy’ problem \citep[for reviews]{2021SSRv..217...18S}.
One of the possibilities is that $k_\mathrm{bol}$ is underestimated and $\eta$ is overestimated for estimating $\Delta M$ in TDEs. This is because at the early phase of TDEs, the accretion is
dominated by super-Eddington accretion flow, in which
the emission is suggested to be dominated in the unobservable extreme UV band
rather in X-ray band \citep{2018ApJ...859L..20D,2019MNRAS.483..565C,2025MNRAS.540.3069P,2025MNRAS.539.3473Q}. Meanwhile, $\eta$ also cannot be well constrained from the observation since both of the bolometric luminosity and the mass accretion rate cannot be estimated accurately. In general, one expect that $\eta$ is below 0.1 for the super-Eddington accretion, which has been discussed in many theoretical and simulation studies  \citep{1978MNRAS.184...53B,1988ApJ...332..646A,1999ApJ...516..420W,2000PASJ...52..133W,2005ApJ...628..368O,2011ApJ...736....2O,2014ApJ...796..106J,2014MNRAS.439..503S,2014MNRAS.441.3177M,2016MNRAS.456.3929S}. The previously used value of $\eta=0.1$ will also cause the underestimation of $\Delta M$.

In \citet{2025MNRAS.539.3473Q} (hereafter \citetalias{2025MNRAS.539.3473Q}), the authors performed the radiation hydrodynamic simulation under a more realistic environment of TDEs, i.e. injecting a continuous mass flow rate at the circularization radius in the form of $\dot M_\mathrm{inject}\propto t^{-5/3}$ for the mass supply rate. The authors further post-processed the simulation data to calculate the emergent spectra by using the method of Monte Carlo radiation transfer. It is found that the emission is indeed dominated in extreme UV at the early phase of the super-Eddington accretion flow in TDEs. In \citetalias{2025MNRAS.539.3473Q}, one group of standard simulation was performed, in which the BH mass was fixed at $M_\mathrm{BH}=10^6M_\odot$.

In this paper, following \citetalias{2025MNRAS.539.3473Q}, we perform three groups of radiation hydrodynamic simulations to study the accretion process of TDEs by taking  $M_{\mathrm{BH}}=10^6,\ 10^{6.5}$ and $10^7M_\odot$ respectively. We select $t=16$ day after the injection of matter at the 
circularization radius as a typical time to post-process the accretion flow and  calculate the emergent spectra. 
Based on the spectra, we integrate the emission in X-ray band and the total spectra to calculate the X-ray luminosity and the bolometric luminosity respectively, and further
calculate the value of $k_\mathrm{bol}$ accordingly. We plot $k_\mathrm{bol}$ as a function of $\theta$ for different $M_\mathrm{BH}$. We found that for TDEs with typical $M_\mathrm{BH}$ in the range of $10^6-10^7M_\odot$, $k_\mathrm{bol}$ is substantially greater than unity across all viewing angles. 
Meanwhile, we also calculate the radiative efficiency $\eta$, showing that $\eta$ is smaller than the typically adopted value of 0.1. In this paper, we take $t=16$ day as typical time for the emergent spectra to calculate $k_\mathrm{bol}$ and $\eta$. This is reasonable because the accreted mass has filled the gap between the BH and the circularization radius, and the stable super-Eddington accretion is ongoing at 16 day in all the three groups of simulations. The detailed discussion of this point can be found in the second paragraph of Section \ref{Summary}.

This paper is organized as follows. In Section \ref{Numerical method}, we describe the numerical methods, including the setup for the radiation-hydrodynamic simulations performed with the ATHENA++ and the Monte Carlo radiative transfer calculations. The simulation results and emergent spectra are presented in Section \ref{Simulation results} and \ref{Spectrum results} respectively. The bolometric correction factor, radiative efficiency and the application for explaining the `missing energy' problem are discussed in Section \ref{Bolometric correction}. Summary and discussions are shown in Section \ref{Summary}.

\begin{figure*}
    \centering
    \begin{subfigure}{0.32\textwidth}
        \centering
        \includegraphics[width=\linewidth]{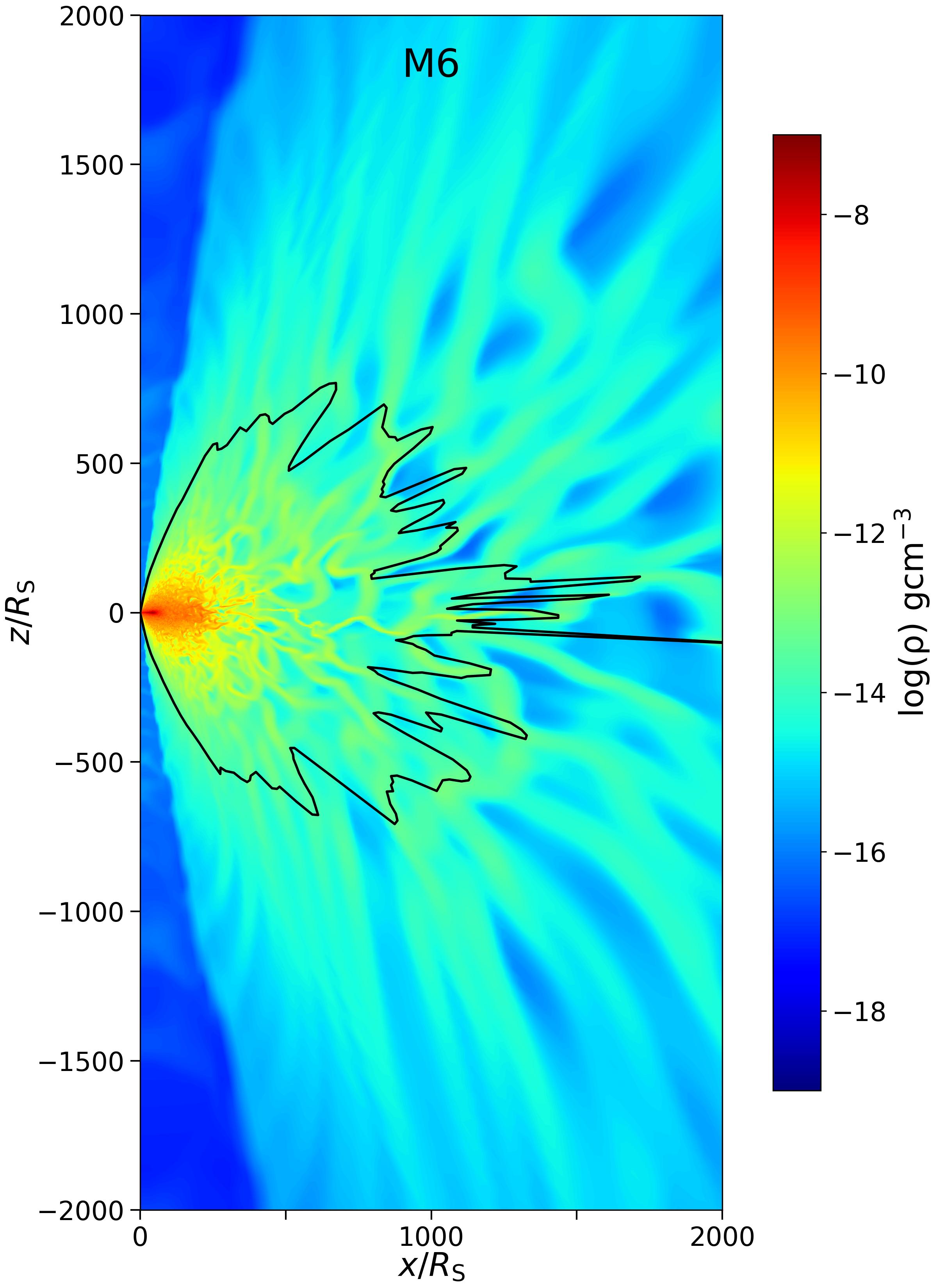}
    \end{subfigure}
    \begin{subfigure}{0.32\textwidth}
        \centering
        \includegraphics[width=\linewidth]{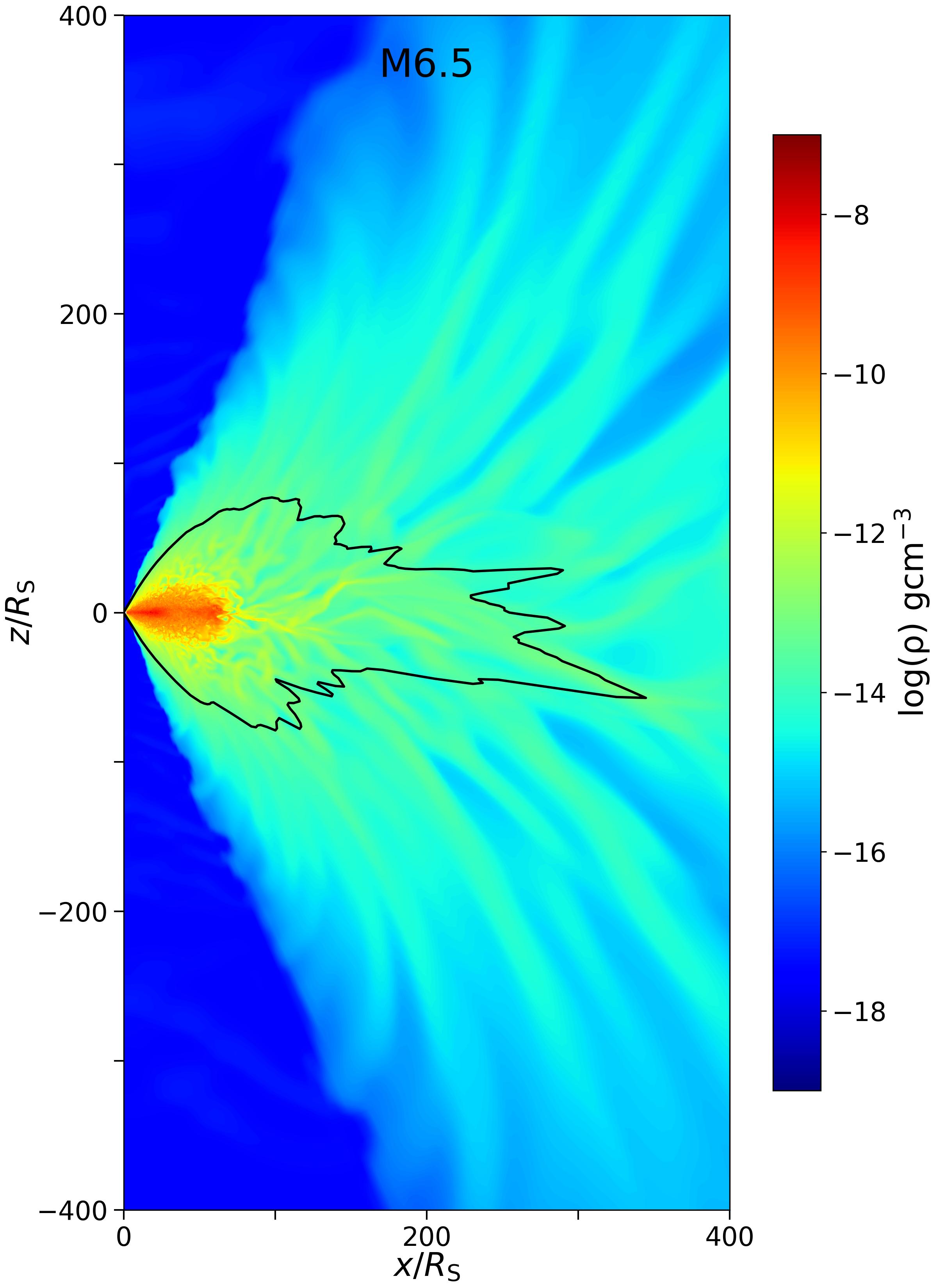}
    \end{subfigure}
    \begin{subfigure}{0.32\textwidth}
        \centering
        \includegraphics[width=\linewidth]{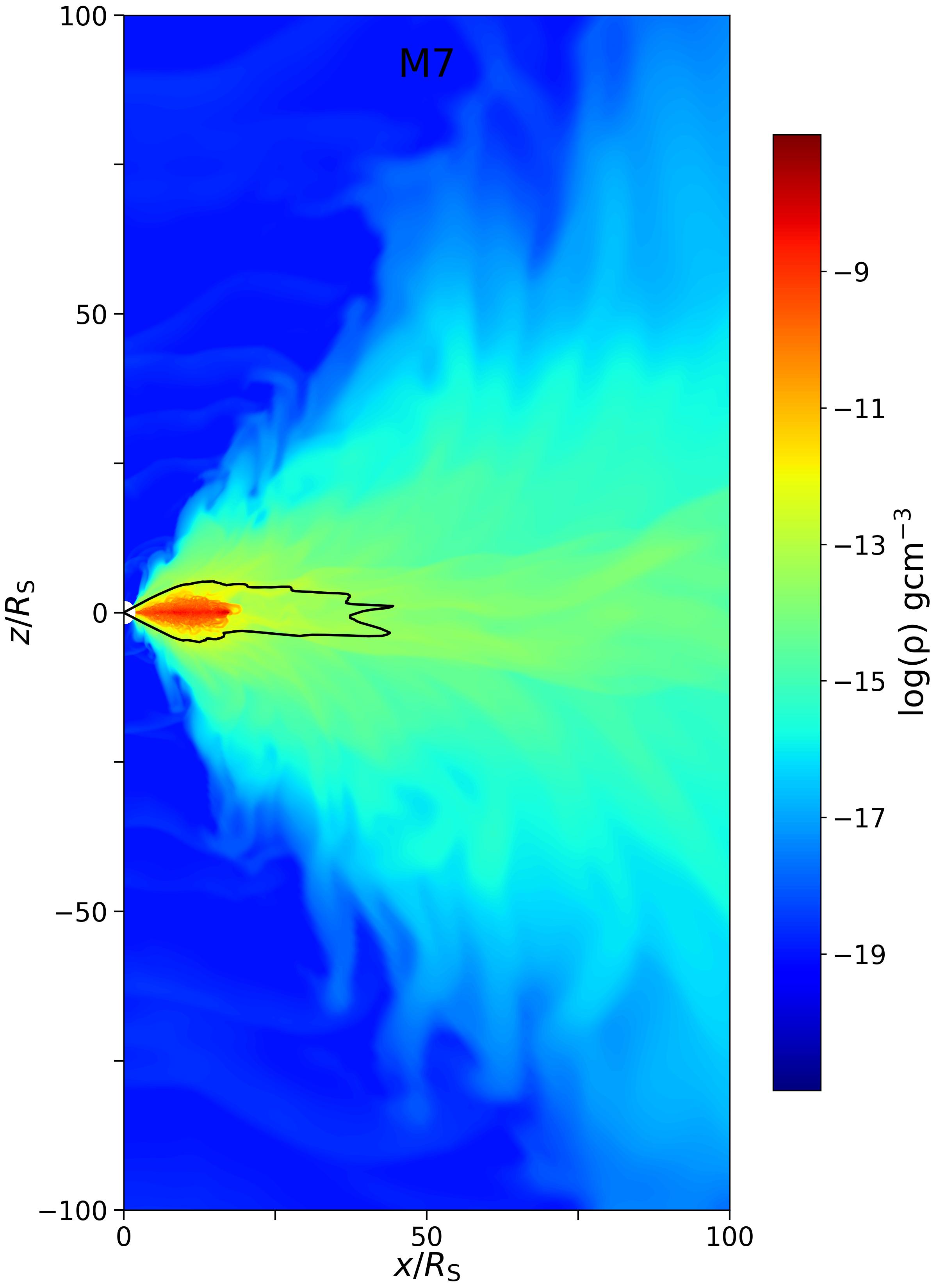}
    \end{subfigure}
    \caption{Snapshots of gas density (in $r-\theta$ plane) at $t=16$ day since the injection of matter at the circularization radius for the simulations of M6, M6.5 and M7 respectively. The black thick line indicates the photosphere of electron scattering.}
	\label{fig1}
\end{figure*}

\section{NUMERICAL METHOD}\label{Numerical method}
We performed two-dimensional axisymmetric radiation hydrodynamic simulations by using the ATHENA++ code \citep{2020ApJS..249....4S,athena} to obtain the properties of super-Eddington accretion flow around the BH in the scenario of TDE. We adopt the spherical coordinates
($r$, $\theta$, $\phi$) in the simulations. The equations of hydrodynamics coupled with the time dependent radiative transfer equations to be solved were described in detail in \citetalias{2025MNRAS.539.3473Q}. We performed three groups of simulations, i.e., M6, M6.5 and M7, in which 
a solar-type star (with mass $M_{*}=M_\odot$ and radius $R_*=R_\odot$) with a parabolic orbit being disrupted by a BH with the mass of $10^6$, $10^{6.5}$ and $10^7M_\odot$.

We used the pseudo-Newtonian potential to mimic the effects of general relativity around a Schwarzschild BH, i.e.,
\begin{equation}
    \phi=-\frac{GM_\mathrm{BH}}{r-R_\mathrm{S}}, 
\end{equation}
where $G$ is the gravitational constant and $R_\mathrm{S}=2GM_\mathrm{BH}/c^2$ is Schwarzschild radius.

To replicate the accretion process in a more realistic environment of TDE, we followed the setup in \citetalias{2025MNRAS.539.3473Q} by continuously injecting mass near the equatorial plane along the $\phi$ direction at circularization radius $R_\mathrm{c}$ = 2$R_\mathrm{t}$, where $R_\mathrm{t}=R_*(M_\mathrm{BH}/M_*)^{1/3}$ is the tidal disruption radius. $R_\mathrm{c}$ is approximately 47$R_\mathrm{S}$, 22$R_\mathrm{S}$, and 10$R_\mathrm{S}$ for the simulations of M6, M6.5, and M7. The mass injection rate is $\dot{M}_{\mathrm{inject}}(t)=\dot{M}_{\mathrm{fb}}(t)=\frac{1}{3}(M_*/t_{\mathrm{fb}})(1+t/t_{\mathrm{fb}})^{-5/3}$, where $t_{\mathrm{fb}}$ is the fallback timescale. $t$=0 is defined as the time when the most bound debris circularized, the corresponding peak fallback rate is $133.8\dot{M}_{\mathrm{Edd}}$, $23.8\dot{M}_{\mathrm{Edd}}$ and $4.2\dot{M}_{\mathrm{Edd}}$ for the simulations of M6, M6.5 and M7 respectively, where $\dot{M}_{\mathrm{Edd}}=L_\mathrm{Edd}/\eta c^2=1.39\times 10^{18}(M_{\mathrm{BH}}/{M_\odot})\ {\rm g~ s^{-1}}$ with radiative efficiency $\eta=0.1$ assumed. $L_\mathrm{Edd}$ is the Eddington luminosity.
%We adopted identical computational domains and grid configurations in our simulations. 
For all the three simulations, the computational domain is $2R_\mathrm{S}\leq r \leq10^5R_\mathrm{S}$ in radial direction and is $0\leq \theta \leq\pi$ in $\theta$ direction. The resolution of the computational domain is set to $N_\mathrm{r}\times N_\mathrm{\theta} = 768\times 256$. Outﬂow boundary conditions are set at the inner and outer radial boundary, while at $\theta=0$ and $\theta=\pi$ we use the reflecting boundary conditions. For more details of the simulations, one can refer to \citetalias{2025MNRAS.539.3473Q}.

Based on the simulation results, we post-processed the accretion flow and calculated the emergent spectra. We utilized the Monte Carlo radiation transfer to compute the emergent spectra for different viewing angles $\theta$ of the super-Eddington accretion flow of TDEs with PYTHON code \citep{2002ApJ...579..725L}, which has been applied to calculate radiative transfer in outflows across various accretion systems, from cataclysmic variables to active galactic nuclei \citep{2013MNRAS.436.1390H,2014ApJ...789...19H,2015MNRAS.450.3331M,2016MNRAS.458..293M}. We selected $t=16$ day at the early phase of TDE, by which time the stable accretion flow had been formed in all three simulations. We took the polar coordinate system, in consistent with the axisymmetric spherical polar coordinate used in the simulations. The density and velocity of the wind as well as the initial temperature were input into the calculation following the result of hydrodynamic simulations.
PYTHON code will then generate the spectrum by tracing the photon propagation in 3D after recalculating the temperature and ionization state until a stable state is reached by iterative calculations.

We took the radius where the $\theta$-direction averaged velocity field transits from outflow to inflow as the inner boundary, which was approximately equal to the radius of the mass injection point. The outer boundary was placed at $\sim 10000R_\mathrm{S}$. We set a central radiation source at the inner boundary. 
The photons from the inner boundary in every iteration were assumed to be a single blackbody with a temperature around $6\times 10^5$ K as in 
\citetalias{2025MNRAS.539.3473Q}.
There is a little difference for setting this blackbody temperature in the present paper. In \citetalias{2025MNRAS.539.3473Q}, the blackbody temperature
was calculated by averaging the radiation temperature along $\theta-$direction at the inner boundary. Instead, we selected the radiation temperature at the transition point of inflow and outflow along the $\theta$-direction at the inner boundary as the temperature of the central blackbody, which was approximately 2-5 $\times \mathrm{10^5\ K}$ for the three groups of simulations in the present paper. This setting method made the calculation for the emission from the inner boundary more accurate.

\begin{figure*}
    \centering

    \begin{subfigure}{0.6\textwidth}
        \centering
        \includegraphics[width=\linewidth]{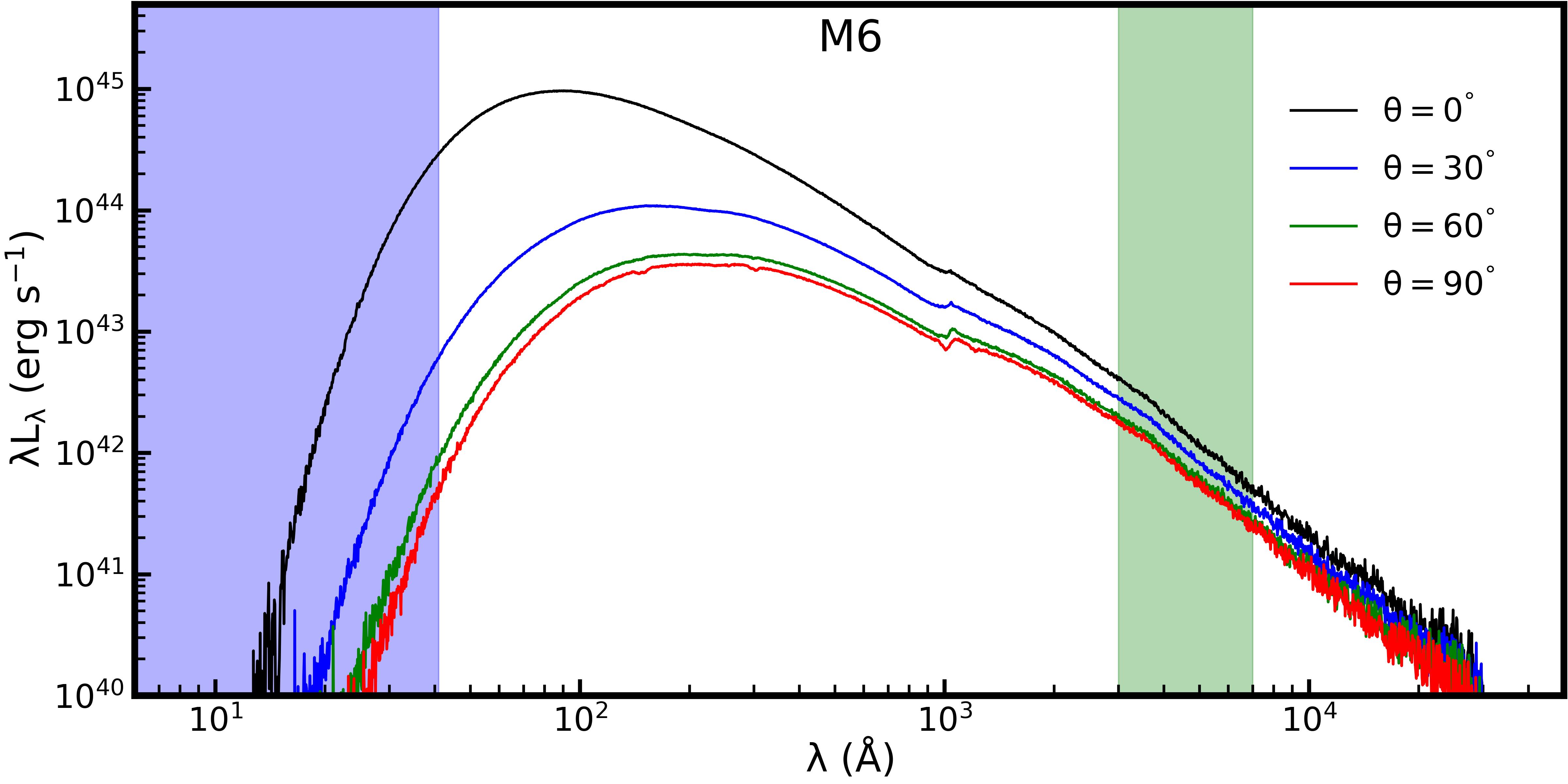}
    \end{subfigure}

    \begin{subfigure}{0.6\textwidth}
        \centering
        \includegraphics[width=\linewidth]{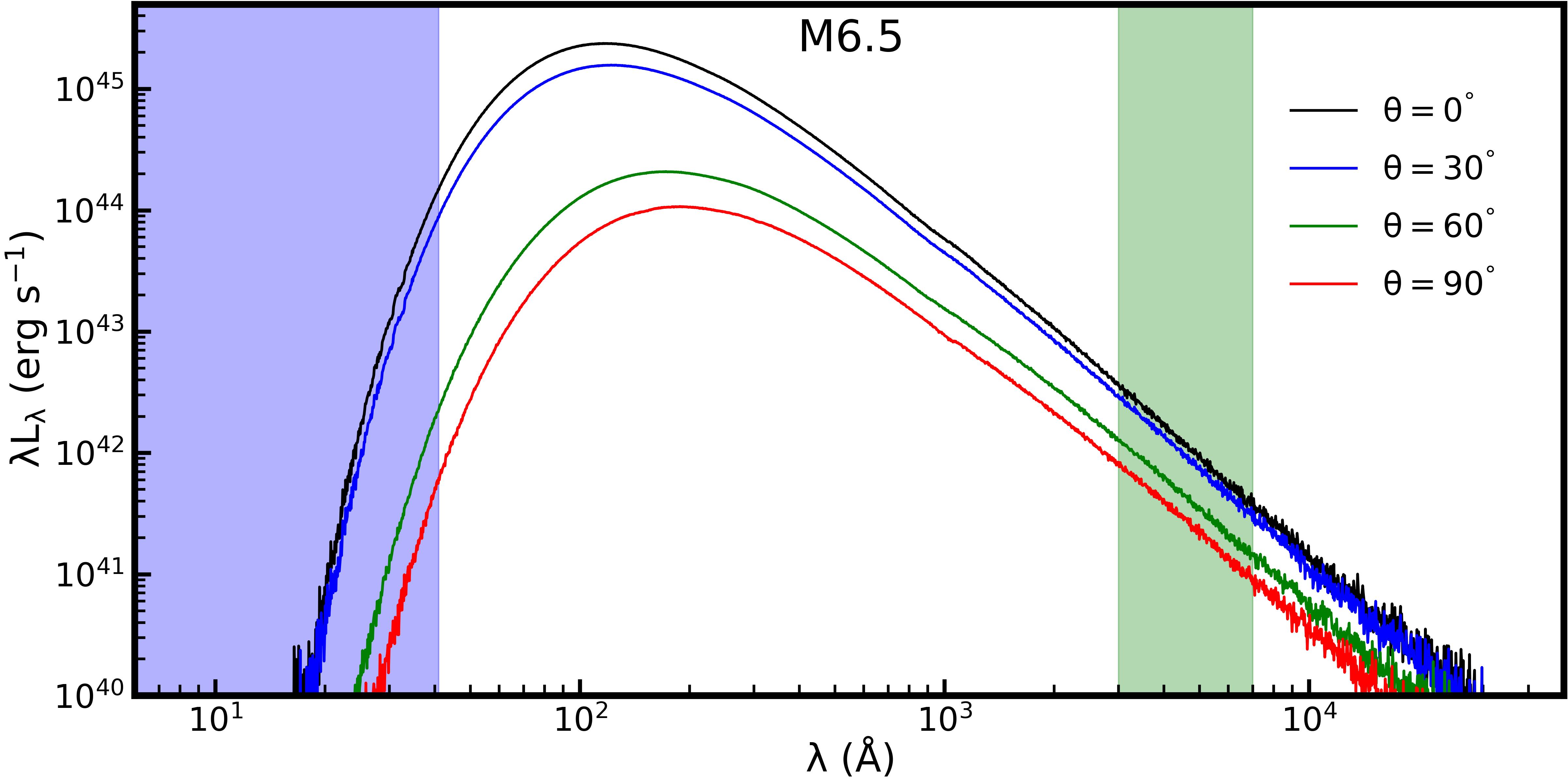}
    \end{subfigure}

    \begin{subfigure}{0.6\textwidth}
        \centering
        \includegraphics[width=\linewidth]{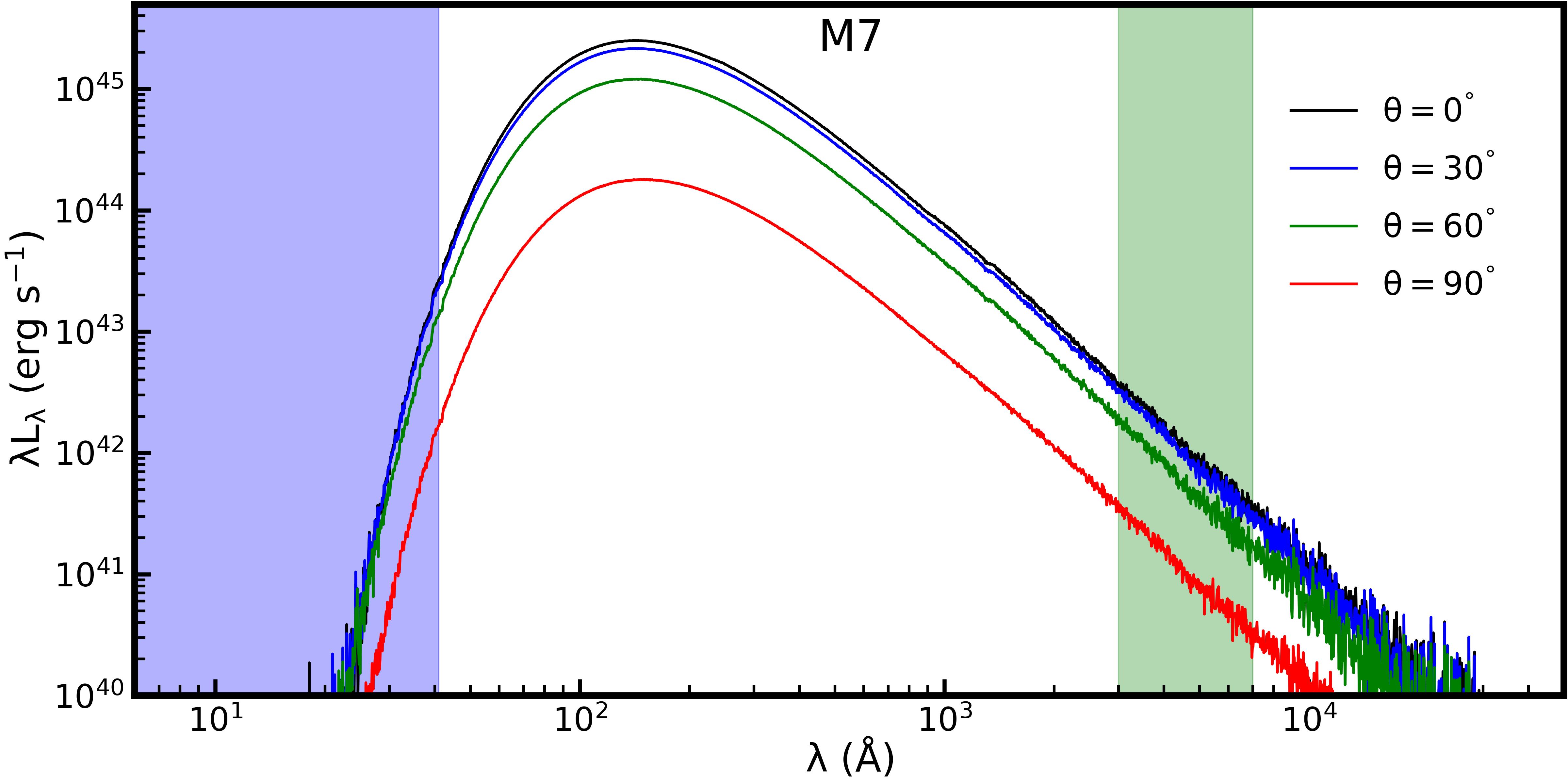}
    \end{subfigure}
    \caption{Emergent spectra of different viewing angles $\theta$ at $t=16$ day since the injection of matter at the circularization radius for the simulations of M6, M6.5 and M7 respectively. The purple shaded region indicates X-ray band of 0.3-2~keV and the green shaded region indicates optical band of 3000-7000$\text{\AA}$.}
    \label{fig2}
\end{figure*}

\section{RESULTS}
\subsection{Results of radiation hydrodynamic simulation}\label{Simulation results}
In Figure \ref{fig1}, we plot the snapshots of the density structure of the accretion inflow and outflow for the simulations of M6, M6.5 and M7 respectively at $t=16$ day after the initiation of gas injection. 
The left panel shows the results of M6 simulation. The injected gas has filled the region between the injection point and BH. In this case a stable accretion flow has been formed. The outflow has extended to several thousand $R_\mathrm{S}$. The gas density reaches its maximum in the grid cells near the equatorial plane, decreases with both decreasing $\theta$ and increasing radius.

We defined the electron scattering photosphere as the location where the optical depth $\tau_\mathrm{\theta,r}=-\int_{10^5R_\mathrm{S}}^{r}\rho \kappa_\mathrm{es}\mathrm{d}r'$ equals to unity along each $\theta$, which is indicated by the black thick curve in Figure \ref{fig1}. $\kappa_\mathrm{es}=0.34$ is the electron scattering opacity assuming solar abundance. When $\theta$ is small, i.e. at the near face-on orientation, the observer can directly observe the inner region, where the X-ray emission is produced. As $\theta$ increases, X-ray emission from the inner region is obscured by the optically thick outflow.

In the middle and right panels, a stable accretion flow has also been formed at $t=16$ day for the simulations of M6.5 and M7.
Meanwhile, it can be seen that 
both the radial extent (in unit of $R_\mathrm{S}$) and thickness of the accretion flow, as well as photosphere decreases for simulation with higher $M_\mathrm{BH}$, indicates relatively stronger accretion flow is formed in the simulation with smaller $M_\mathrm{BH}$. The trend of the density as a function of $\theta$ is similar to that of M6, which can be reflected in the emergent spectra as will be shown in the next section.

\subsection{Emergent spectrum}\label{Spectrum results}
Based on the numerical simulations, we computed the emergent spectra at $t=16$ day. 
In Figure \ref{fig2}, we plot the emergent spectra of four different viewing angles, i.e.,  $0^\mathrm{o},\ 30^\mathrm{o},\ 60^\mathrm{o}$ and $90^\mathrm{o}$ for $M_{\mathrm{BH}}=10^{6},\ 10^{6.5}$ and $10^7M_\odot$ respectively. In general, it can be seen from Figure \ref{fig2} that 
the emission is dominated by extreme UV band for $M_{\mathrm{BH}}=10^{6},\ 10^{6.5}$ and $10^7 M_\odot$ respectively, and the multi-band luminosity decreases with increasing $\theta$ as has been shown in \citet{2019MNRAS.483..565C} and \citet{2025MNRAS.539.3473Q}. 

From the top panel of Figure \ref{fig2}, it can be seen that for M6 the bolometric luminosity decreases about 20 times, while X-ray luminosity drops roughly three orders of magnitude for
viewing angle $\theta$ increasing from $0^\mathrm{o}-90^\mathrm{o}$. This can be understood as follows. In general, the X-ray emission is from the innermost region of the accretion flow. For a lower $\theta$, the X-ray emission can directly escape to be observed, while for a higher $\theta$, the X-ray emission is reprocessed via scattering, free–free, bound–free, and bound–bound absorption in the optically thick outflow, and re-emits in the optical and UV bands. As for the decrease of bolometric luminosity, it is dominated by the decrease of UV luminosity. 
In the super-Eddington accretion, the UV emission is dominated by the photosphere. The decrease of UV luminosity is related to several factors
of the properties of the photosphere, such as the solid angle of the accretion flow seen by the distant observer, the distribution of the radiation temperature in the photosphere, and the velocity of the grid cell at the photosphere etc. In general, since the shape of the photosphere is closer to 
being spherical, the change of the bolometric luminosity is milder than that of 
the X-ray luminosity.

As for M6.5 and M7, the trend of the X-ray luminosity and the bolometric luminosity decrease with the increasing $\theta$ is similar to that of M6. However, it is clear that the variation magnitude of the X-ray luminosity with $\theta$ is smaller compared with the M6 simulation. 
This is because the Eddington scaled mass injection rate $\dot m_{\rm inject}$ (defined as $\dot M_{\rm inject}/\dot M_{\rm Edd}$) for M6.5 and M7 is lower than M6. In general, a lower $\dot m_{\rm inject}$ will predict a relatively weak outflow in the previous simulation as also shown in Figure \ref{fig1} in the present paper. In this case, the shape of the outflow is significantly deviated from spherical shape, which will lead to the X-ray emission to escape to be observed in a wider viewing angle respect to the polar direction rather than be reprocessed in the optically thick outflow. On the other hand, the spherical shaped outflow will lead to an isotropic photosphere and optical emission, therefore the optical emission shows a more significant viewing-angle dependence in the simulations of M6.5 and M7 than in the simulation of M6.

\begin{figure*}
    \centering
    \begin{minipage}{0.49\textwidth}
        \centering
        \includegraphics[width=1\linewidth]{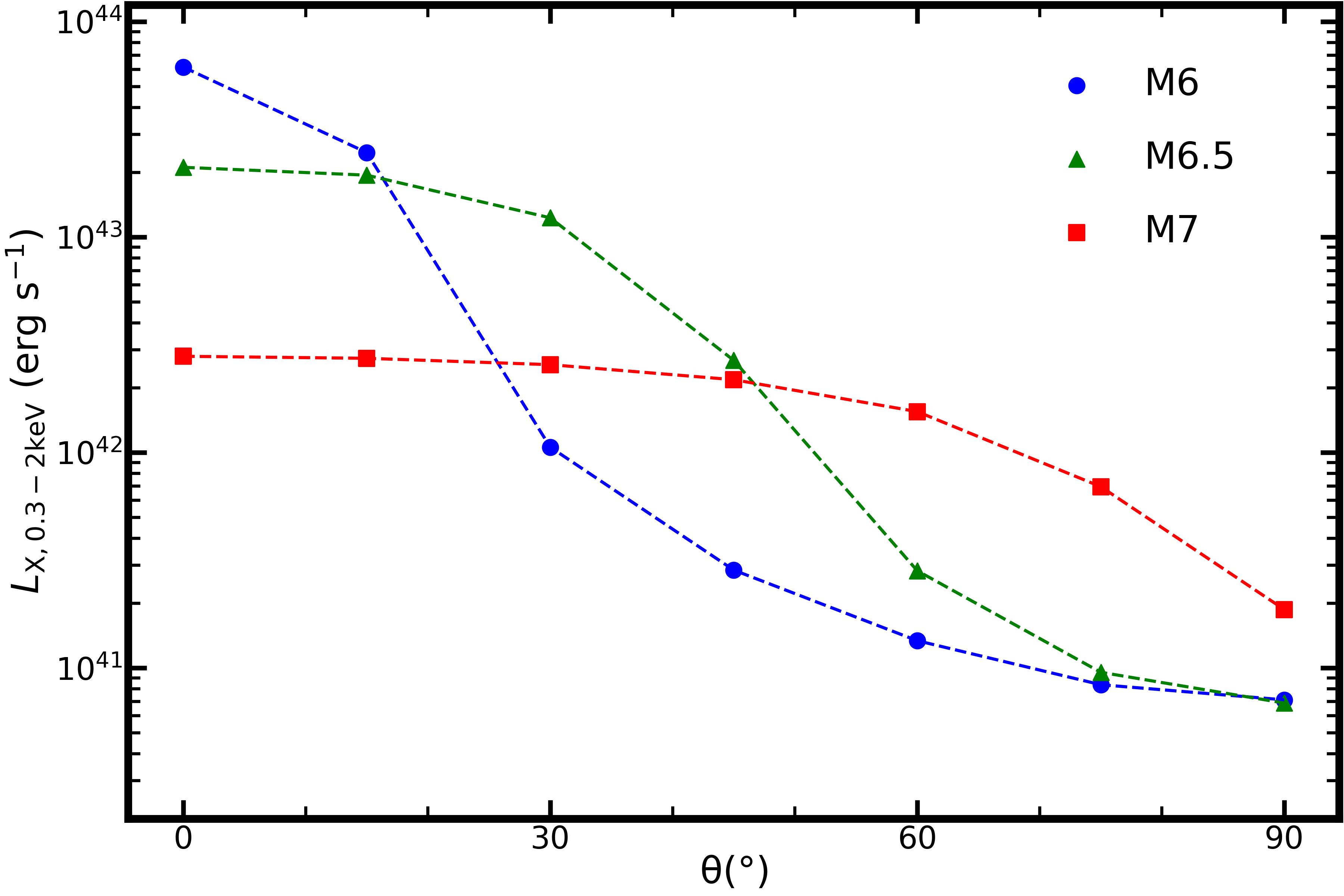}
    \end{minipage}
    \hfill 
    \begin{minipage}{0.49\textwidth}
        \centering
        \includegraphics[width=1\linewidth]{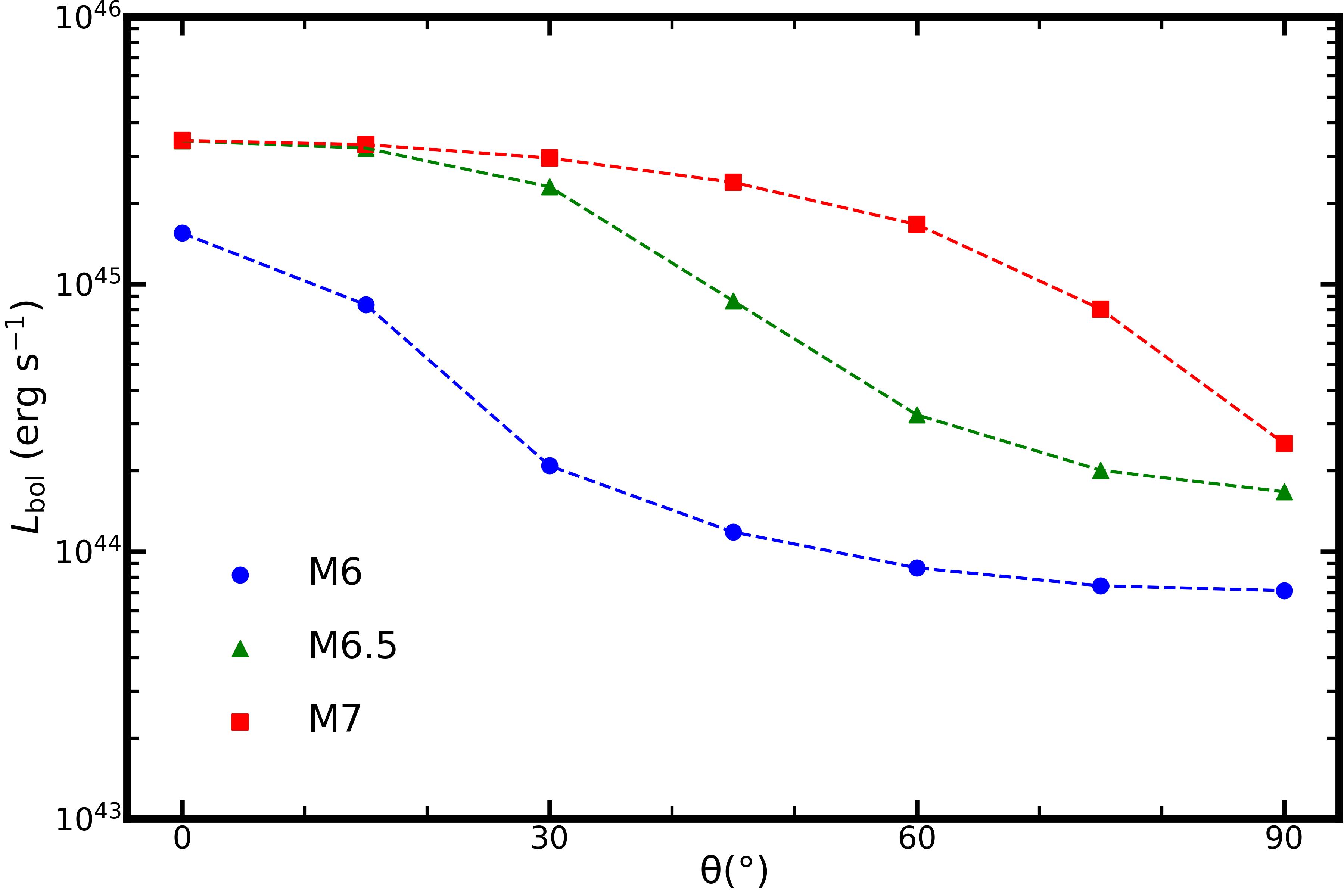}
    \end{minipage}
    \caption{Left panel: X-ray luminosity $L_\mathrm{X,\ 0.3-2keV}$ as a function of $\theta$ at $t=16$ day since the injection of matter at the circularization radius for the simulations of M6, M6.5, and M7 respectively. Right panel: Bolometric luminosity $L_\mathrm{bol}$ as a function of $\theta$ at $t=16$ day since the injection of matter at the circularization radius for the simulations of M6, M6.5 and M7 respectively. Both of $L_\mathrm{X,\ 0.3-2keV}$ and $L_\mathrm{bol}$ represent the isotropic-equivalent luminosities.}
    \label{fig3}
\end{figure*}

We define X-ray band as 0.3-2~keV, which is commonly used in many studies of TDEs \citep[e.g.,][for review]{2021ARA&A..59...21G}. To more clearly show the X‑ray and bolometric luminosities as a function of $\theta$ for the simulations of M6, M6.5 and M7, we integrate the spectra over 0.3–2~keV and over the wavelength range used in the radiation transfer calculation which is 4-30000~\AA\ to derive the X‑ray luminosity $L_\mathrm{X,\ 0.3-2keV}$ and the bolometric luminosity $L_\mathrm{bol}$ respectively. 
Here both $L_\mathrm{X,\ 0.3-2keV}$ and $L_\mathrm{bol}$ are angle-dependent luminosities.
In the left panel of Figure \ref{fig3}, we plot $L_\mathrm{X,\ 0.3-2keV}$ as a function of $\theta$ for simulations of M6, M6.5 and M7 respectively.
In M6 simulation, $L_\mathrm{X,\ 0.3-2keV}$ decreases from about $7\times10^{43}$ erg/s at $0^\mathrm{o}$ to roughly $7\times10^{40}$ erg/s at $90^\mathrm{o}$, corresponding to a decrease by a factor of $\sim$1000. For M6.5, $L_\mathrm{X,\ 0.3-2keV}$ decreases from $2\times10^{43}$ erg/s to $7\times10^{40}$ erg/s, by a factor of $\sim300$. While in M7 simulation, $L_\mathrm{X,\ 0.3-2keV}$ decreases from $3\times10^{42}$ erg/s to $2\times10^{41}$ erg/s, showing only about one order of magnitude decrease. From the spectra, $L_\mathrm{X,\ 0.3-2keV}$ is in the range of about $7\times10^{40}-7\times10^{43}$ erg/s, which covers the typical luminosity of the non-jetted X-ray TDEs.
In the right panel of Figure \ref{fig3}, we plot $L_\mathrm{bol}$ as a function of $\theta$ for simulations of M6, M6.5 and M7 respectively. 
In the M6 simulation, $L_\mathrm{bol}$ decreases from $1.4\times10^{45}$ erg/s to $7\times10^{43}$ erg/s for $\theta$ in the range of $0^\mathrm{o}-90^\mathrm{o}$, while for the M6.5 and M7 simulations, it varies from $3.3\times10^{45}$ erg/s to $1.6\times10^{44}$ erg/s and from $3.3\times10^{45}$ erg/s to $2.3\times10^{44}$ erg/s respectively. For all the three simulations, $L_\mathrm{bol}$ shows about 15-20 times drop. 

\begin{figure*}
    \centering

    \begin{subfigure}{0.49\textwidth}
        \centering
        \includegraphics[width=1\linewidth]{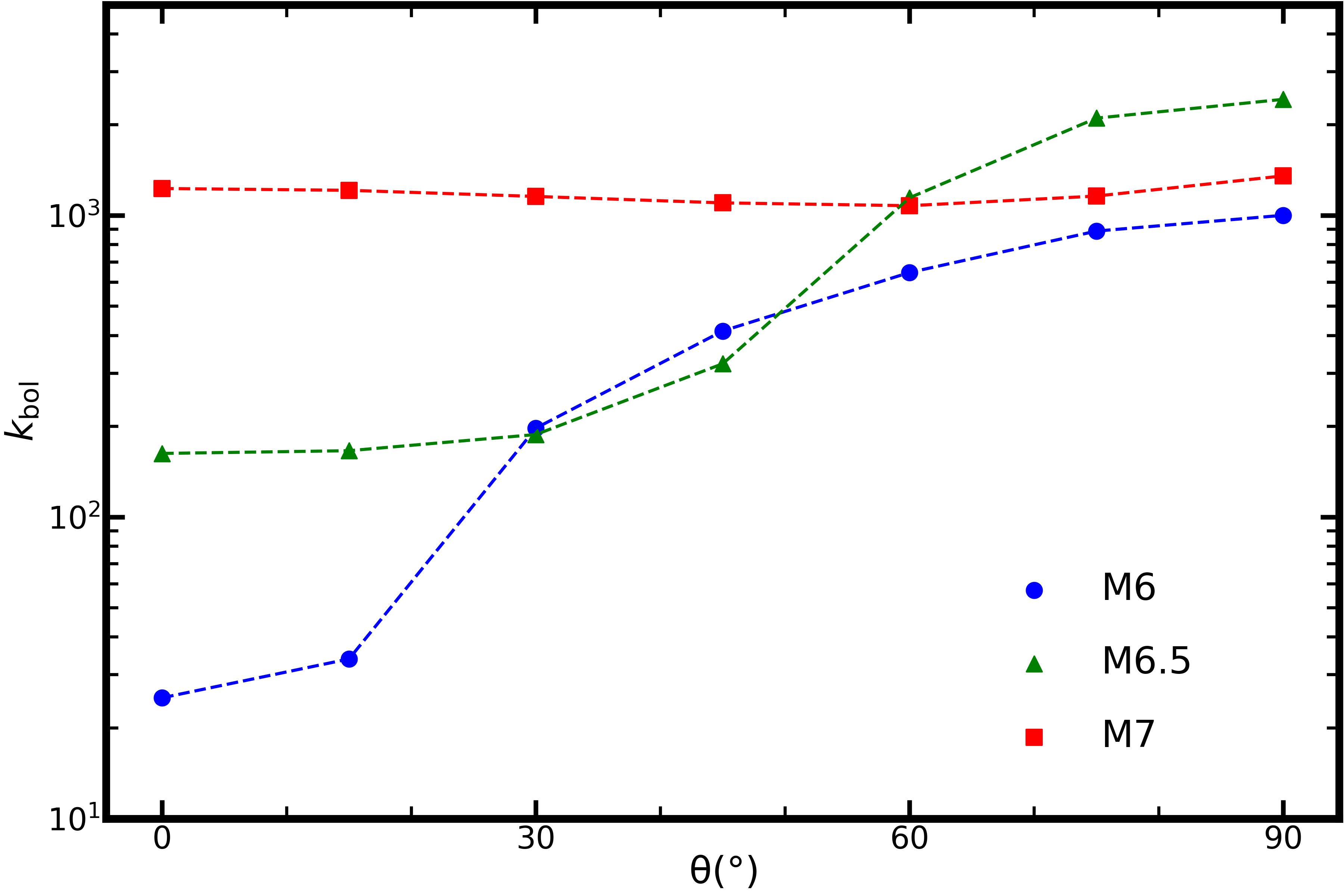}
    \end{subfigure}
    \begin{subfigure}{0.49\textwidth}
        \centering
        \includegraphics[width=\linewidth]{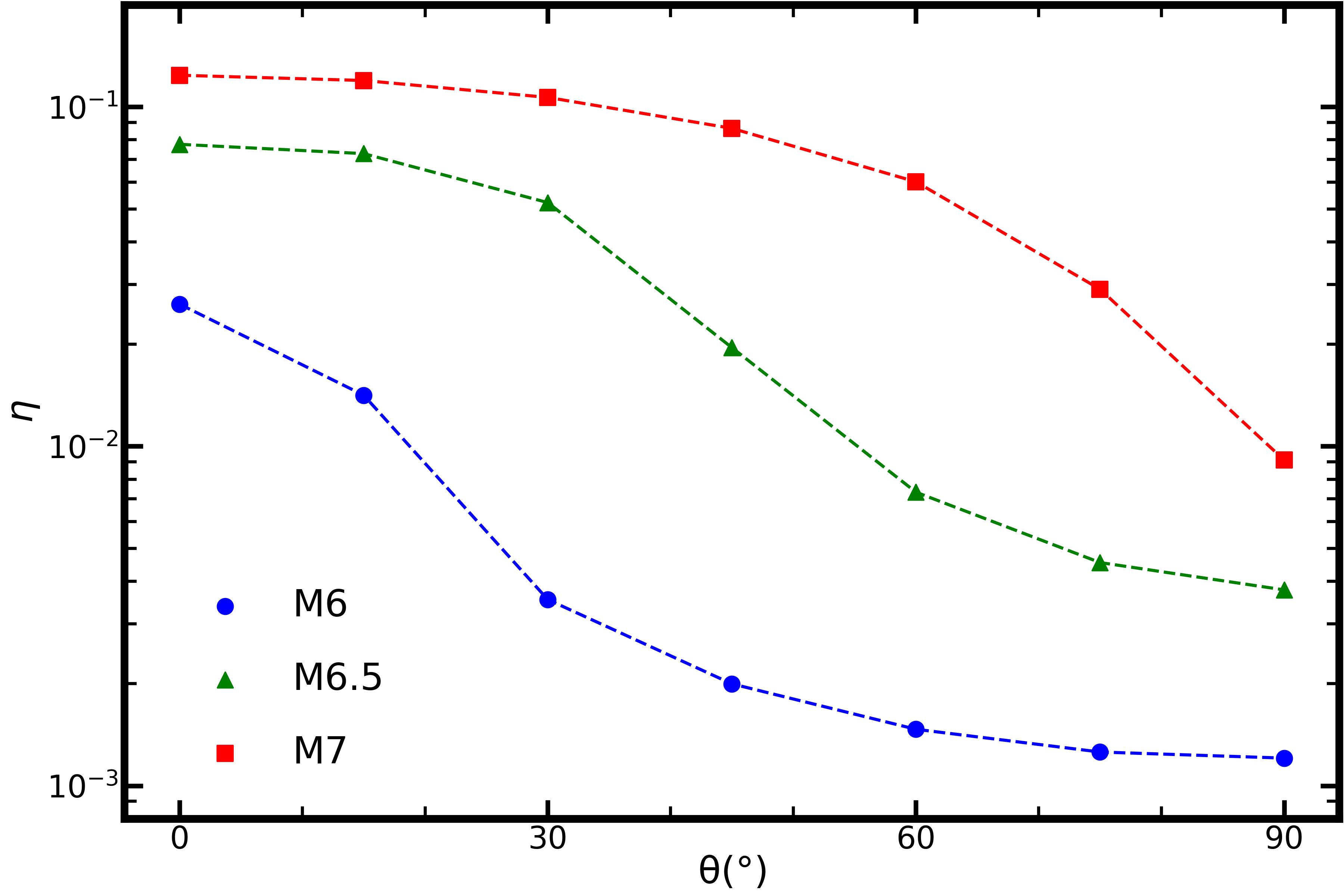}
    \end{subfigure}
    \caption{Left panel: X-ray bolometric correction factor $k_\mathrm{bol}$ as a function of $\theta$
    for the simulations of M6, M6.5, and M7 respectively. Right panel: Radiative efficiency $\eta$ derived from the emergent spectra as a function of $\theta$ for the simulations of M6, M6.5 and M7 respectively. Both of $k_\mathrm{bol}$ and $\eta$ are derived from the isotropic-equivalent luminosities for each $\theta$.}
    \label{fig4}
\end{figure*}

\begin{figure*}
    \centering
    \begin{subfigure}{0.49\textwidth}
        \centering
        \includegraphics[width=1\linewidth]{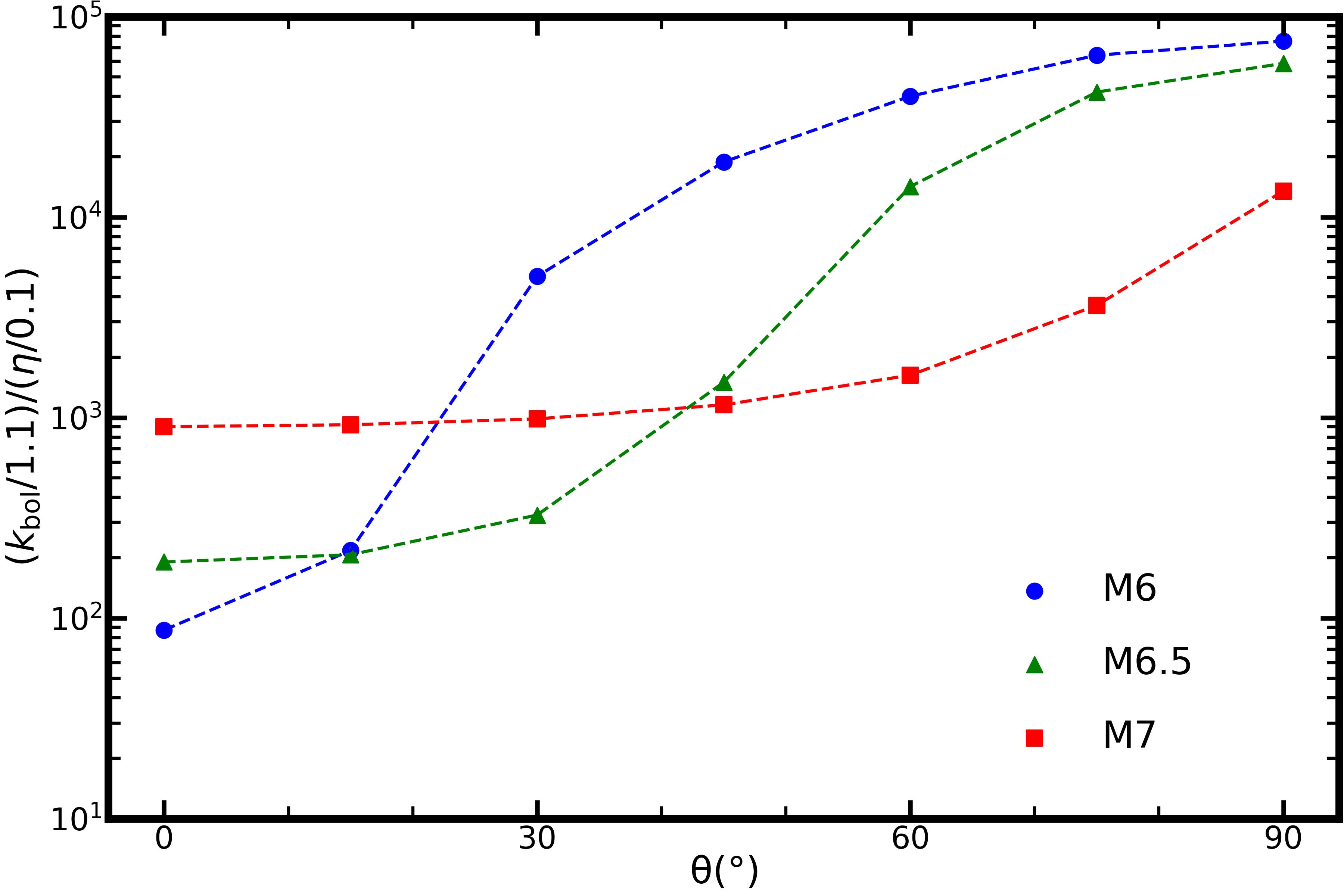}
    \end{subfigure}
    \caption{The ratio of $k_\mathrm{bol}/\eta$ to the typical values used of 1.1/0.1 as a function of $\theta$ for the simulations of M6, M6.5 and M7 respectively. $k_\mathrm{bol}/\eta$ is derived from the isotropic-equivalent luminosities for each $\theta$.}
    \label{fig5}
\end{figure*}

\subsection{X-ray bolometric correction factor and radiative efficiency}\label{Bolometric correction}
In the left panel of Figure \ref{fig4}, we plot the X-ray bolometric correction factor $k_\mathrm{bol}$ as a function of $\theta$ for the simulations of M6, M6.5 and M7 respectively. Here $k_\mathrm{bol}$ is defined as 
\begin{equation}
    k_\mathrm{bol}=\frac{L_\mathrm{bol}}{L_\mathrm{X,\ 0.3-2keV}}. 
    \label{eq2}
\end{equation}
Here $k_\mathrm{bol}$ represents the correction factor between the isotropic-equivalent X-ray luminosity and bolometric luminosity for each $\theta$. 
In the M6 simulation, $k_\mathrm{bol}$ increases from 25 to 1000 as $\theta$ increases from $0^\mathrm{o}$ to $90^\mathrm{o}$, varying by a factor of 40. For M6.5 simulation, $k_\mathrm{bol}$ increases from 200 to 2000, increases by a factor of 10. For the M7 simulation, $k_\mathrm{bol}$ remains nearly constant at about 1000 across all $\theta$. It can be seen that $k_\mathrm{bol}$ ranges from several tens to several thousands assuming different $\theta$ and $M_\mathrm{BH}$, which is substantially larger than the commonly adopted value of unity for bolometric correction.

The radiative efficiency $\eta$ is plotted in the right panel of Figure \ref{fig4}, which is calculated as 
\begin{equation}
    \eta=\frac{L_\mathrm{bol}}{\dot{M}_\mathrm{in}c^2}.
    \label{eq3}
\end{equation}
$\dot{M}_\mathrm{in}$ is the accretion rate at the inner boundary, defined as $\dot{M}_\mathrm{in}=2\times 2\pi (2R_\mathrm{S})^2 \int_0^{\pi}\rho \min(0,v_\mathrm{r})\sin\theta \mathrm{d}\theta$, where $\rho$ and $v_\mathrm{r}$ is the density and radial velocity of the gas at the inner boundary $2R_\mathrm{S}$ respectively.
We should note that $L_\mathrm{bol}$ is angle-dependent bolometric luminosity, therefore
$\eta$ here is the apparent radiative efficiency inferred by an observer at a given inclination angle rather than the intrinsic radiative efficiency of the accreting system.
From the right panel of Figure \ref{fig4}, one can see $\eta$ is in the range of 0.001–0.02, 0.004-0.08 and 0.01–0.13 for simulations of M6, M6.5 and M7 respectively and decreases with increasing $\theta$ in simulations, since $L_\mathrm{bol}$ decreases with increasing $\theta$ as in Figure \ref{fig2}.
We also note that $\eta$ is systematically higher for a higher $M_\mathrm{BH}$. This is because at $t=16$ day, 
$\dot M_{\rm in}=47.3\dot{M}_\mathrm{Edd}$ for M6, 
$\dot M_{\rm in}=11.2\dot{M}_\mathrm{Edd}$ for M6.5 and $\dot M_{\rm in}=2.2\dot{M}_\mathrm{Edd}$ for M7. The increase of $\dot M_{\rm in}$ will result in the decrease of $\eta$, which is consistent with our general understanding that radiative efficiency will decrease with increasing mass accretion rate for super-Eddington accretion flow due to the effect of the advection \citep{1978MNRAS.184...53B,1988ApJ...332..646A,2005ApJ...628..368O,2011ApJ...736....2O,2014MNRAS.439..503S,2016MNRAS.456.3929S}. 
It can be seen that the derived $\eta$ in most cases from simulations (except for $\theta\leq30^\mathrm{o}$ in M7) is lower than the commonly adopted value of 0.1.

In Figure \ref{fig5}, we plot the ratio of $k_\mathrm{bol}/\eta$ to the typically used value 1.1/0.1 as a function of $\theta$ for simulations of M6, M6.5 and M7 respectively. For M6 simulation, this ratio is in the range of $90-8\times10^4$, and for M6.5 and M7 simulations this ratio is in the range of $200-6\times10^4$ and $900-1\times10^4$ respectively. The value of $k_\mathrm{bol}/\eta$ increases with increasing $\theta$ for all the three simulations. It can also be seen that, $k_\mathrm{bol}/\eta$ crosses a wider range of value for $\theta$ increasing from $0^\mathrm{o}$ to $90^\mathrm{o}$ for simulations with lower $M_\mathrm{BH}$.
These derived ranges indicate that when a solar‑type star is tidally disrupted by a BH with mass of $10^6-10^7M_\odot$, $k_\mathrm{bol}/\eta$ is roughly 2–5 orders of magnitude larger than the commonly adopted value of 1.1/0.1, therefore the accreted mass $\Delta M$ is also roughly 2–5 orders of magnitude larger than the values inferred from previous estimations \citep{2002ApJ...576..753L,2008A&A...489..543E,2009A&A...495L...9C,2010ApJ...722.1035M,2017A&A...598A..29S}. To present our results more clearly for better application, we list the value of $L_\mathrm{X,\ 0.3-2keV}$, $L_\mathrm{bol}$, $k_\mathrm{bol}$, $\eta$ and $k_\mathrm{bol}/\eta$ for each selected $M_\mathrm{BH}$ and $\theta$ in Table \ref{table1}. 
In the following, we will show that if the derived value in the present paper is used, $\Delta M$ can be corrected to a reasonable range comparable with $0.5M_\odot$.

\begin{table*}
	\centering
    \caption{The isotropic-equivalent value of $L_\mathrm{X,0.3-2keV}$, $L_\mathrm{bol}$, $k_\mathrm{bol}$, $\eta$ and $k_\mathrm{bol}/\eta$ for each selected $M_\mathrm{BH}$ and $\theta$.}
	\begin{tabular}{ccccccc} 
		\hline
		  $M_\mathrm{BH}$  & $\theta$ & $L_\mathrm{X,\ 0.3-2keV}~\mathrm{(erg~s^{-1})}$ & $L_\mathrm{bol}~\mathrm{(erg~s^{-1})}$ & $k_\mathrm{bol}$ & $\eta$ & $k_\mathrm{bol}/\eta$ \\
		\hline
        $10^6~M_\mathrm{\odot}$ & $0^\mathrm{o}$ & $6.16\times10^{43}$ & $1.55\times10^{45}$ & 25.16 & $2.62\times10^{-2}$ & $9.59\times10^{2}$ \\
         & $15^\mathrm{o}$ & $2.47\times10^{43}$ & $8.36\times10^{44}$ & 33.85 & $1.41\times10^{-2}$ & $2.39\times10^{3}$ \\
         & $30^\mathrm{o}$ & $1.06\times10^{42}$ & $2.09\times10^{44}$ & 197.17 & $3.54\times10^{-3}$ & $5.58\times10^{4}$ \\
         & $45^\mathrm{o}$ & $2.85\times10^{41}$ & $1.18\times10^{44}$ & 414.03 & $2.00\times10^{-3}$ & $2.07\times10^{5}$ \\
         & $60^\mathrm{o}$ & $1.34\times10^{41}$ & $8.67\times10^{43}$ & 647.01 & $1.47\times10^{-3}$ & $4.41\times10^{5}$ \\
         & $75^\mathrm{o}$ & $8.37\times10^{40}$ & $7.43\times10^{43}$ & 887.69 & $1.26\times10^{-3}$ & $7.06\times10^{5}$ \\
         & $90^\mathrm{o}$ & $7.12\times10^{40}$ & $7.13\times10^{43}$ & 1001.40 & $1.21\times10^{-3}$ & $8.30\times10^{5}$ \\
        \hline
        $10^{6.5}~M_\mathrm{\odot}$ & $0^\mathrm{o}$ & $2.11\times10^{43}$ & $3.43\times10^{45}$ & 162.56 & $7.76\times10^{-2}$ & $2.10\times10^{3}$ \\
         & $15^\mathrm{o}$ & $1.94\times10^{43}$ & $3.22\times10^{45}$ & 165.98 & $7.28\times10^{-2}$ & $2.28\times10^{3}$ \\
         & $30^\mathrm{o}$ & $1.23\times10^{43}$ & $2.31\times10^{45}$ & 187.80 & $5.22\times10^{-2}$ & $3.60\times10^{3}$ \\
         & $45^\mathrm{o}$ & $2.68\times10^{42}$ & $8.64\times10^{44}$ & 322.39 & $1.95\times10^{-2}$ & $1.65\times10^{4}$ \\
         & $60^\mathrm{o}$ & $2.83\times10^{41}$ & $3.24\times10^{44}$ & 1144.88 & $7.33\times10^{-3}$ & $1.56\times10^{5}$ \\
         & $75^\mathrm{o}$ & $9.56\times10^{40}$ & $2.01\times10^{44}$ & 2102.51 & $4.55\times10^{-3}$ & $4.63\times10^{5}$ \\
         & $90^\mathrm{o}$ & $6.88\times10^{40}$ & $1.67\times10^{44}$ & 2427.33 & $3.78\times10^{-3}$ & $6.43\times10^{5}$ \\
        \hline
        $10^7~M_\mathrm{\odot}$ & $0^\mathrm{o}$ & $2.80\times10^{42}$ & $3.44\times10^{45}$ & 1228.57 & $1.24\times10^{-1}$ & $9.92\times10^{3}$ \\
         & $15^\mathrm{o}$ & $2.74\times10^{42}$ & $3.32\times10^{45}$ & 1211.68 & $1.20\times10^{-1}$ & $1.01\times10^{4}$ \\
         & $30^\mathrm{o}$ & $2.56\times10^{42}$ & $2.96\times10^{45}$ & 1156.25 & $1.07\times10^{-1}$ & $1.08\times10^{4}$ \\
         & $45^\mathrm{o}$ & $2.18\times10^{42}$ & $2.40\times10^{45}$ & 1100.92 & $8.64\times10^{-2}$ & $1.27\times10^{4}$ \\
         & $60^\mathrm{o}$ & $1.55\times10^{42}$ & $1.67\times10^{45}$ & 1077.42 & $6.01\times10^{-2}$ & $1.79\times10^{4}$ \\
         & $75^\mathrm{o}$ & $6.95\times10^{41}$ & $8.06\times10^{44}$ & 1159.71 & $2.90\times10^{-2}$ & $3.99\times10^{4}$ \\
         & $90^\mathrm{o}$ & $1.87\times10^{41}$ & $2.53\times10^{44}$ & 1352.94 & $9.11\times10^{-3}$ & $1.48\times10^{5}$ \\
        \hline
	\end{tabular}

    \label{table1}
\end{table*}

With the derived $k_\mathrm{bol}/\eta$, we can correct the total accretion mass $\Delta M$ for the X-ray TDEs with the assumption of $M_\mathrm{BH}$ in the range of $10^6-10^7M_\odot$ as examples. 
For NGC 5905, the authors derived the released X-ray energy $\Delta E_\mathrm{X}$ was $4.5\times 10^{49}$ erg by integrating the X-ray luminosity over time, while the mass of the central BH was estimated to be $\sim 10^7M_\odot$ by using the relation of \citet{2000MNRAS.317..488S} for spiral galaxies \citep{2002RvMA...15...27K}. Assuming $M_\mathrm{BH}=10^7M_\odot$, we therefore calculated the value of $k_\mathrm{bol}/\eta$, which ranges from $1\times10^{4}$ to $1.8\times10^{4}$ for $\theta$ between $0^\mathrm{o}$ and $60^\mathrm{o}$. We further derived that $\Delta M$ to be in the range of $0.25-0.45M_\odot$, which is comparable with the expected value of $\Delta M\sim0.5M_\odot$ for disrupting a solar-type star.
For another TDE identified in the ROSAT archive, TDXF1347-3254, $M_\mathrm{BH}$ was estimated to be $10^7M_\odot$ and $\Delta E_\mathrm{X}\sim2\times 10^{50}$ erg. Thus the estimated $\Delta M$ is in the range of $1.1-2M_\odot$, which is slightly larger than $\sim0.5M_\odot$ and may imply that a more massive star was disrupted. 

For NGC 3599 and SDSS J132341.97+482701.3 discussed in \citet{2008A&A...489..543E}, $M_\mathrm{BH}$ was estimated to be $1.3(\pm 0.6)\times10^6M_\odot$ and $2.2(\pm 0.9)\times10^6M_\odot$ by using the $M_\mathrm{BH}-\sigma$ relationship from \citet{2005SSRv..116..523F}, and to be $3.1(\pm 1.0)\times10^6M_\odot$ and $5.0(\pm 1.4)\times10^6M_\odot$ by using the $M_\mathrm{BH}-\sigma$ estimate from \citet{2007ApJ...664..226L}, respectively. The $M_\mathrm{BH}$ of both NGC 3599 and SDSS J132341.97+482701.3 are roughly in the range of $10^{6-6.5}M_\odot$. Within this $M_\mathrm{BH}$ range, for $\theta=0^\mathrm{o},\ 30^\mathrm{o}$ and $60^\mathrm{o}$, the value of $k_\mathrm{bol}/\eta$ is in the range of $1\times10^{3}-2\times10^{3}$, $3.6\times10^{3}-5.6\times10^{4}$ and $1.5\times10^{5}-4.4\times10^{5}$, respectively. For NGC 3599, $\Delta E_\mathrm{X}$ is calculated to be $7.1\times10^{48}$ erg, thus the inferred $\Delta M$ is in the range of $3\times10^{-3}-8\times10^{-3}M_\odot$, $0.015-0.22M_\odot$ and $0.6-1.7M_\odot$ for $\theta=0^\mathrm{o},\ 30^\mathrm{o}$ and $60^\mathrm{o}$ respectively. For SDSS J132341.97+482701.3, $\Delta E_\mathrm{X}$ was calculated to be $7.6\times10^{50}$ erg. At $\theta=0^\mathrm{o}$, $\Delta M$ ranges from $0.4M_\odot$ to $0.9M_\odot$, while at $\theta=15^\mathrm{o}$, $k_\mathrm{bol}/\eta$ is about $2.4\times10^{3}$ for both M6 and M6.5, $\Delta M \sim 1M_\odot$. For $\theta\geq 15^\mathrm{o}$, the derived $\Delta M$ is significantly larger than $1M_\odot$. Therefore, for SDSS J132341.97+482701.3, $\Delta M$ is roughly comparable with $0.5M_\odot$ if the observer is located very close to the polar direction. While for NGC 3599, if the observer is located at $\theta$ of $30^\mathrm{o}-60^\mathrm{o}$, the derived $\Delta M$ with $k_\mathrm{bol}$ and $\eta$ given in the present paper is also approximately equal to $0.5M_\odot$.

For SDSS J131122.15-012345.6 and 2MASX 0740-85, the estimated $M_\mathrm{BH}$ is $1-7\times10^{6}M_\odot$ and $3.5_{-2.4}^{+6.5}\times10^{6}M_\odot$ respectively. Thus we calculate $\Delta M$ by assuming $M_\mathrm{BH}$ is in the range of $10^{6-7}M_\odot$. For SDSS J131122.15-012345.6, $\Delta E_\mathrm{X}$ is $\sim 1.7\times10^{50}$ erg, thus $\Delta M$ is in the range of $0.1-1M_\odot$ at $\theta=0^\mathrm{o}$, and in the range of $0.3-5.3M_\odot$ at $\theta=30^\mathrm{o}$ for $M_\mathrm{BH}$ in the range of $10^{6-7}M_\odot$. For 2MASX 0740-85, $\Delta E_\mathrm{X} \sim 5\times10^{50}$ erg. $\Delta M$ is in the range of $0.27-2.8M_\odot$ at $\theta=0^\mathrm{o}$, and in the range of $1-15M_\odot$ at $\theta=30^\mathrm{o}$ for $M_\mathrm{BH}$ in the range of $10^{6-7}M_\odot$.
For ASASSN-14li the inferred $M_\mathrm{BH}$ equals to $2\times10^6M_\odot$ \citep{2015Natur.526..542M}, and $\Delta E_\mathrm{X} \sim 3.5\times10^{50}$ erg. Thus we consider the $M_\mathrm{BH}$ in the range of $10^{6-6.5}M_\odot$. For $M_\mathrm{BH}=10^{6}M_\odot$, $\Delta M$ is $0.18M_\odot$ and $0.47M_\odot$ at $\theta=0^\mathrm{o}$ and $\theta=15^\mathrm{o}$ respectively and for $M_\mathrm{BH}=10^{6.5}M_\odot$, $\Delta M$ is $0.4M_\odot$, $0.44M_\odot$ and $0.7M_\odot$ at $\theta=0^\mathrm{o}$, $\theta=15^\mathrm{o}$ and $\theta=30^\mathrm{o}$ respectively.

From the X-ray TDEs we discussed above one can see that in most cases where the derived $\Delta M$ is comparable to $0.5M_\odot$ for $\theta\lesssim30^\mathrm{o}$. According to the unified model of TDEs, X‑ray selected TDEs are typically observed at relatively small inclinations \citep{2018ApJ...859L..20D}, and our estimations are consistent with this picture.

\section{SUMMARY AND DISCUSSION}\label{Summary}
In this work, we performed radiation‑hydrodynamic simulations using the ATHENA++ code to study the accretion process following the tidal disruption of a solar‑type star on a parabolic orbit by BHs of $M_{\mathrm{BH}}=10^{6},\ 10^{6.5}$ and $10^7 M_\odot$ respectively. We selected $t=16$ day since the injection of matter at the circularization radius as a typical time to show the spectral features of 
super-Eddington accretion flow in the early phase of TDE accretion. Specifically, we conducted post‑processing radiation transfer calculations using Monte Carlo method to compute the emergent spectra. 
Based on the emergent spectra, we calculated isotropic-equivalent $k_\mathrm{bol}$ and $\eta$ within equation (\ref{eq2}) and equation (\ref{eq3}) respectively.
We plotted $(k_\mathrm{bol}/1.1)/(\eta/0.1)$ as functions of $\theta$ for different BH masses in Figure \ref{fig5}. It was found that $(k_\mathrm{bol}/1.1)/(\eta/0.1)$ is both dependent on BH mass $M_\mathrm{BH}$ and viewing angle $\theta$. Specifically, for $M_\mathrm{BH}$ in the range of $10^{6-7}\ M_\odot$, $(k_\mathrm{bol}/1.1)/(\eta/0.1)$ is in the range of $\sim90-8\times10^4$ for $\theta$ from $0^\mathrm{o}-90^\mathrm{o}$.
We further calculated the accreted mass $\Delta M$ for X-ray TDE NGC 5905, TDXF1347-3254, NGC 3599, SDSS J132341.97+482701.3, SDSS J131122.15-012345.6, 2MASX 0740-85 and ASASSN-14li during their outburst by choosing appropriate $M_\mathrm{BH}$ and $\theta$ for $k_\mathrm{bol}$ and $\eta$. We found that $\Delta M$ for all these TDEs can be corrected to a reasonable value of $\sim 0.5 M_\odot$ as expected for a solar-type main sequence star disrupted by a supermassive BH.

We argue that the proposed correction can alleviate the missing energy problem, but such solution strongly depend on $\theta$, which can not be accurately determined from observations. Therefore our solution for correcting the value of $k_\mathrm{bol}$ and $\eta$ is only one possible explanation for the missing energy problem.
Furthermore, our solution for making $\Delta M\sim 0.5 M_\odot$ mostly require small inclination angles. This is consistent with the so called unified model of TDEs, in which the X-ray emission is expected to be observed in a smaller viewing angle. 

%most of our calculations which gives $\Delta M\sim 0.5 M_\odot$ are in the small inclination, where the X-ray emission is expected according to the unified model of TDEs. This will support that our results are reasonable in solving this problem.}
%Due to such reason, we need to clarify that the corrections to $k_\mathrm{bol}$ and $\eta$ presented in this work represent only one possible explanation for the missing energy problem. 
% If the inclination angle is different from the range we select, a different value of $\Delta M$ will be given.

Here we should note that in the present paper we take $t=16$ day as a typical time to calculate the emergent spectra, and then calculate the bolometric correction factor $k_\mathrm{bol}$ and the radiative efficiency $\eta$. 
However, as we know in TDE environment, the fallback rate decreases with time. This will result in the mass injection rate decreases with time as implemented in the simulations of the present paper. 
In general, with the decrease of the mass injection rate, the properties of the accretion flow will change, including the strength of the mass inflow rate, outflow rate and the velocity of the outflow etc, all of which will influence the emergent spectra of the accretion flow. 
However, we expect that such a change of the properties of the accretion flow and the corresponding emergent spectra will be minor since the accretion flow of a typical TDE can keep in super-Eddington phase for relatively longer timescales. The dynamics and the radiation of super-Eddington accretion flow are relatively stable as has been shown in many previous studies \citep{2016MNRAS.456.3929S,2018PASJ...70..108K,2019ApJ...880...67J,2022ApJ...937L..28T,2025MNRAS.539.3473Q}.
In the present paper, we take $M_\mathrm{BH}=10^6,\ 10^{6.5}$ and $10^7M_\odot$ for the simulations. For $M_\mathrm{BH}=10^6M_\odot$, the initial mass injection $\dot{M}_\mathrm{inject} = 133.8\dot{M}_\mathrm{Edd}$. In this case, there will be roughly $730$ day that the accretion flow
can be evolved to Eddington accretion rate $\dot M_\mathrm{Edd}$. For $M_\mathrm{BH}=10^{6.5}$ and $10^{7} M_\odot$, this time will be roughly $400$ and $180$ day respectively. So $k_\mathrm{bol}$ will not change much in the timescale of a few hundred days, while 
$\eta$ will have some uncertainties which needs 
more detailed calculations. In addition, observationally, the total
released X-ray energy $\Delta E_\mathrm{X}$ is calculated by integrating the X-ray light curve in the timescale of years. However, at the timescale of years, the accretion rate has dropped below $\dot{M}_\mathrm{Edd}$. In this case the contribution of the accretion to $\Delta M$ is minor. 
So the calculated $\Delta M $ is dominated by the super-Eddington accretion phase, and the derived $k_\mathrm{bol}$ and $\eta$ in the present paper are good approximations.

Besides, in this paper we only considered a solar-type star disrupted by a BH with different masses and set the mass injecting radius at the circularization radius. 
However, TDE is a complex system. 
The property of the disrupted star, penetration factor and circularization efficiency etc, can all affect the fallback rate. Further, this will influence the properties of the accretion flow, and consequently affect the emergent spectra as well as the calculation of $k_\mathrm{bol}$ and $\eta$. 
The effect of these parameters are not tested in this paper, which is very necessary to be done in the future. 
%Therefore, $k_\mathrm{bol}$ and $\eta$ given here are more accurate compared to the previously adopted value, but are not universal values among all TDEs with different parameters.
Here we should also note that the pseudo-Newtonian potential is adopted for mimicking the effects of general relativity around a Schwarzschild BH in the simulation. A more detailed treatment of relativistic effects may affect the simulation results. The spin of BH as well as the magnetic field are also not considered in this paper. All of these will have potential influence on the properties of the accretion flow and the corresponding value of $k_\mathrm{bol}$ and $\eta$. 
To more accurately constrain $k_\mathrm{bol}$ and $\eta$ in the long-term evolution and in a lager parameter space in the environment of TDEs, further simulations and spectra calculations are needed in the future, which however exceeds the research in the present paper. 

Bolometric correction is also an important question in active galactic nuclei (AGNs). The bolometric correction for AGNs has been studied in some previous works, in which the 2-10~keV hard X-ray luminosity are generally used to estimate the bolometric luminosity. It is found that 2-10~keV correction factor is positive correlated with the Eddington ratio. The 2-10~keV correction factor increases from $\sim 10$ to several hundreds as the Eddington ratio increases from about 0.1 to 10 \citep[e.g.,][]{2012MNRAS.420.1825J,2012MNRAS.425..907J}. The hard X-ray emission around a BH is expected to originate from the corona \citep[][for review]{2002ApJ...572L.173L,2003ApJ...587..571L,2012ApJ...744..145Q,2013ApJ...764....2Q,2017MNRAS.467..898Q,2018MNRAS.477..210Q,2022iSci...25j3544L}, which, however, is difficult to generate in our simulations. Therefore in this paper, we use the 0.3-2~keV soft X-ray luminosity to correct the bolometric luminosity. We should note that in our simulation the 0.3-2~keV soft X-ray emission is from the accretion disk. While in the scenario of AGNs, the origin of soft X-ray emission is not fully determined, which could be from the transition layer between the accretion disk and the corona \citep{1998MNRAS.301..179M,2011A&A...534A..39M,2012MNRAS.420.1848D,2013A&A...549A..73P}, or from the relativistically-smeared reflection component of the accretion disk \citep{1993MNRAS.261...74R,2005MNRAS.358..211R,2006MNRAS.365.1067C,2009Natur.459..540F,2013MNRAS.428.2901W}.

\section*{Acknowledgements}
We thank the very useful discussions with Wei Chen and Chenlei Guo from NAOC, and the participants of the TDE FORUM (Full-process Orbital to Radiative Unified Modeling) online seminar series for their inspiring discussions.
This work is supported by the Strategic Priority Research Program of the Chinese Academy of Science (Grant No.XDB0550200), Shandong Provincial Key Research and Development Program (No.2022CXGC020106), National Super-computing Center in Jinan for computing resource, National Key R\&D Program of China (No.2023YFA1607903) and National Natural Science Foundation of China (Grant No. 12173048, 12333004).

%%%%%%%%%%%%%%%%%%%%%%%%%%%%%%%%%%%%%%%%%%%%%%%%%%
\section*{Data Availability}

The data underlying this article will be shared on reasonable request to the corresponding author.

%%%%%%%%%%%%%%%%%%%% REFERENCES %%%%%%%%%%%%%%%%%%

% The best way to enter references is to use BibTeX:

\bibliographystyle{mnras}
\bibliography{ref} % if your bibtex file is called example.bib

% Alternatively you could enter them by hand, like this:
% This method is tedious and prone to error if you have lots of references
%\begin{thebibliography}{99}
%\bibitem[\protect\citeauthoryear{Author}{2012}]{Author2012}
%Author A.~N., 2013, Journal of Improbable Astronomy, 1, 1
%\bibitem[\protect\citeauthoryear{Others}{2013}]{Others2013}
%Others S., 2012, Journal of Interesting Stuff, 17, 198
%\end{thebibliography}

%%%%%%%%%%%%%%%%%%%%%%%%%%%%%%%%%%%%%%%%%%%%%%%%%%

%%%%%%%%%%%%%%%%% APPENDICES %%%%%%%%%%%%%%%%%%%%%

%\appendix

%\section{Some extra material}

%If you want to present additional material which would interrupt the flow of the main paper,
%it can be placed in an Appendix which appears after the list of references.

%%%%%%%%%%%%%%%%%%%%%%%%%%%%%%%%%%%%%%%%%%%%%%%%%%

% Don't change these lines
\bsp	% typesetting comment
\label{lastpage}
\end{document}